\documentclass{aa}
\usepackage{txfonts}
\usepackage{graphicx}

\usepackage{natbib}
\newcommand{\hess}{H.E.S.S.}
\newcommand{\funit}{10$^{-13}$\,cm$^{-2}$s$^{-1}$TeV$^{-1}$}
\newcommand{\fu}{cm$^{-2}$s$^{-1}$TeV$^{-1}$}
\newcommand{\ifunit}{10$^{-12}$\,cm$^{-2}$s$^{-1}$}
\newcommand{\spectralindex}{$3.09\,\pm\,0.24_\mathrm{stat}\,\pm\,0.10_\mathrm{sys}$}
\newcommand{\normTeV}{$3.00\,\pm\,0.80_\mathrm{stat}\,\pm\,0.31_\mathrm{sys}$}
\newcommand{\normalisation}{(\normTeV)\,$\times$\,\funit}
\newcommand{\chiprob}{0.47}
\newcommand{\nonstr}{${N_{on}}$}
\newcommand{\noffstr}{${N_{off}}$}
\newcommand{\non}{1706}
\newcommand{\noff}{13784}
\newcommand{\alphafactor}{0.0909}
\newcommand{\excess}{453}
\newcommand{\significance}{11.6\,$\sigma$}

\newcommand{\nonT}{4389}
\newcommand{\noffT}{40420}
\newcommand{\excessT}{715}
\newcommand{\significanceT}{10.9\,$\sigma$}
\newcommand{\iflux}{(4.1\,$\pm$\,0.5)\,$\times$\,\ifunit}

\newcommand{\nonWy}{3776}
\newcommand{\noffWy}{35280}
\newcommand{\alphafactorWy}{0.0903}
\newcommand{\excessWy}{591}
\newcommand{\significanceWy}{9.7\,$\sigma$}
\newcommand{\spectralindexWy}{$3.06\,\pm\,0.21_\mathrm{stat}\,\pm\,0.10_\mathrm{sys}$}
\newcommand{\normWy}{$3.07\,\pm\,0.75_\mathrm{stat}\,\pm\,0.31_\mathrm{sys}$}

\newcommand{\chiprobWy}{0.60}

\renewcommand{\aa}{A\&A}
\renewcommand{\apj}{ApJ}

\begin{document}

\title{Discovery of Very High Energy $\gamma$-Ray Emission from the BL Lac Object H\,2356$-$309
	with the \hess\ Cherenkov Telescopes}
\offprints{M. Tluczykont, \email{martin.tluczykont@poly.in2p3.fr}}

\author{F. Aharonian\inst{1}
 \and A.G.~Akhperjanian \inst{2}
 \and A.R.~Bazer-Bachi \inst{3}
 \and M.~Beilicke \inst{4}
 \and W.~Benbow \inst{1}
 \and D.~Berge \inst{1}
 \and K.~Bernl\"ohr \inst{1,5}
 \and C.~Boisson \inst{6}
 \and O.~Bolz \inst{1}
 \and V.~Borrel \inst{3}
% ...
}

\institute{
Max-Planck-Institut f\"ur Kernphysik,
Heidelberg, Germany
\and
 Yerevan Physics Institute, Yerevan,
Armenia
\and
Centre d'Etude Spatiale des Rayonnements, CNRS/UPS, Toulouse, France
\and
Universit\"at Hamburg, Institut f\"ur Experimentalphysik,
Hamburg, Germany
\and
Institut f\"ur Physik, Humboldt-Universit\"at zu Berlin,
Germany
\and
LUTH, UMR 8102 du CNRS, Observatoire de Paris, Section de Meudon,
France
% ...
}

\date{received ... / accepted ...}

\abstract
{
 The extreme synchrotron BL Lac object H\,2356$-$309, located at a redshift of $z = 0.165$, 
 was observed from June to December 2004 
 with a total exposure of $\approx$40\,h live-time
 with the \hess\ (High Energy Stereoscopic System) 
 array of atmospheric-Cherenkov telescopes (ACTs).
 Analysis of this data set 
 {yields, for the first time, a} strong excess of \excess\ $\gamma$-rays
 (10 {standard deviations above background}) from H\,2356$-$309, corresponding to an 
 observed integral flux above 200\,GeV of I($>$200\,GeV) = \iflux\ 
 {(statistical error only)}.
 The differential energy spectrum of the source between
 {200\,GeV} and 1.3\,TeV is well-described by a power law 
 with a normalisation (at 1 TeV) of N$_0$ = \normalisation\ 
 and a photon index of $\Gamma$ = \spectralindex .
 H\,2356$-$309 is one of the most distant BL Lac objects detected at 
 very-high-energy $\gamma$-rays so far.
 Results from simultaneous observations from 
 ROTSE{-III} (optical),
 RXTE (X-rays) 
 and NRT (radio)
 are also included {and used together with the \hess\ data} to constrain 
 a single-zone homogeneous synchrotron self-Compton (SSC) model. 
 {This model provides an adequate fit to the \hess\ data when 
 using a reasonable set of model parameters.}
\keywords{gamma rays: observations -- galaxies: active -- galaxies: BL Lacertae objects: individual: H\,2356$-$309}
}

\authorrunning{Aharonian et al. (the \hess\ Collaboration)}
\titlerunning{Discovery of H\,2356$-$309 \hess}
\maketitle
\section{Introduction}
\label{intro}
The Spectral Energy Distribution (SED) of Active Galactic Nuclei (AGN)
spans the complete electromagnetic spectrum from radio waves
to very-high-energy (VHE; E\,$>$\,100\,GeV) $\gamma$-rays.
In the widely-accepted unified model 
of AGN \citep[e.g.][]{rees:1984a,urry:1995a},
the ``central engine'' of these objects consists
of a super-massive black hole (up to 10$^9$\,M$_\odot$) surrounded
by a thin accretion disk and a dust torus.
In {some} radio-loud AGN, i.e. objects with a radio to B-band flux ratio 
F$_\mathrm{5GHz}$/F$_\mathrm{B}$\,$>$\,10,
two relativistic plasma outflows (jets) {presumably} perpendicular to the 
plane of the accretion disk have been observed.

AGN are known to be VHE $\gamma$-ray emitters 
since the detection of Mrk\,421 above 300\,GeV by the 
Whipple group \citep{punch:1992a}, {{who} pioneered the imaging 
atmospheric-Cherenkov technique.}
At very high energies, a number of AGN ($\approx$10) were subsequently
detected by different groups using a similar technique.
Almost all these objects
{are {BL Lacertae (BL Lac)} objects, {belonging to the
class of Blazars (BL Lac objects and Flat Spectrum Radio Quasars),}}
i.e. AGN having their jet pointing at
a small angle to the line of sight. 
The only confirmed VHE detection of an extragalactic 
object not belonging to the BL Lac class 
is the giant radio galaxy M\,87 \citep{aharonian:2003b,aharonian:2005f}.

Two broad peaks are present in the observed SED of AGN.
The first peak is located in the {radio, optical, and X-ray bands},
the second peak is found at higher energies and can extend to the VHE band.
The observed broad-band emission from AGN is commonly explained
by two different model types.
In leptonic models, the lower-energy peak is explained by synchrotron
emission of relativistic electrons and the high-energy peak 
is assumed to result from inverse Compton (IC) scattering of electrons
off a seed-photon population,
see e.g. \citet[][]{sikora:2001a} and references therein.
In hadronic models, the 
{emission is} assumed to be 
produced via the interactions
of relativistic protons with matter
\citep{pohl:2000a}, ambient photons \citep{mannheim:1993a} 
or magnetic fields \citep{aharonian:2000c}, 
or {via the interactions of relativistic protons with} {photons and magnetic fields} 
\citep{muecke:2001a}.

The observed $\gamma$-ray emission {from BL Lac objects} shows high
{variability}
ranging from short bursts of sub-hour duration
to long-time activity of the order of months.
Detailed studies
of variability of BL Lac type objects can contribute to the understanding of
their intrinsic acceleration mechanisms 
\citep[e.g.][]{krawczynski:2001a,aharonian:2002e}.
Additionally, observations of distant objects
in the VHE band provide an indirect measurement of the
SED of the Extragalactic Background Light 
(EBL), see e.g. \citet{stecker:1992a,primack:1999a} and references therein.
Due to the absorption of VHE $\gamma$-rays via e$^+$e$^-$ pair production
with the photons of the EBL, the shape of the observed VHE
spectra is distorted as compared to the intrinsically emitted spectra.
Using a given spectral shape of the EBL, the observed AGN spectrum 
can be corrected for this absorption.
The resulting intrinsic (i.e., corrected) spectrum can then be
compared to basic model assumptions on the spectral shape of the 
$\gamma$-ray emission, thereby constraining the applied shape of
the EBL.
In this context, it is especially important to detect
AGN at higher redshifts but also to study the spectra of
objects over a wide range of redshifts,
in order to disentangle the effect of the EBL from
the intrinsic spectral shape of the objects.
To date, the redshifts of VHE emitting {BL Lac objects}
{with measured spectra} range from $z = 0.033$ to $z = 0.129$.

The high frequency peaked BL Lac object (HBL) H\,2356$-$309, identified in
the optical by \citet{schwartz:1989a},
is hosted by an elliptical galaxy located at a redshift of $z = 0.165$ \citep{falomo:1991a}.
The object was first detected in X-rays by the
satellite experiment UHURU \citep{forman:1978a}
and subsequently by the Large Area Sky Survey experiment
onboard the HEAO-I satellite \citep{wood:1984a}.
The spectrum of H\,2356$-$309 as 
observed by BeppoSAX \citep{costamante:2001a}
is not compatible
with a single power law model, indicating that the peak of the
synchrotron emission lies within the energy range {of BeppoSAX}.
A broken power law fit yields a synchrotron peak around 1.8\,keV, 
with a detection of the source up to 50\,keV. 
These observations qualified the object as an \emph{extreme synchrotron
blazar}.

A selection of TeV candidate BL Lac objects was proposed by
\citet{costamante:2002a}. 
The objects were selected from
several BL Lac samples and using information in the
radio, optical and X-ray bands. 
VHE predictions for the selected objects were given by the authors
based on a parametrisation proposed by
\citet{fossati:1998a}, suitable for predictions of high state
flux of an average source population.
{The authors also gave VHE flux predictions based on}
a simple one-zone homogeneous
SSC model \citep{ghisellini:2002a},
appropriate for a quiescent state of the 
specific VHE source candidate. 
H\,2356$-$309 is included in this list and 
the predicted integral flux values 
above 300\,GeV for H\,2356$-$309
are 8.4$\times$\ifunit\
for the parametrisation and
1.9$\times$\ifunit\ for the SSC model.
{It should be noted that no absorption due to the EBL was taken into account in these calculations.}

In this paper the discovery of VHE $\gamma$-rays
from H\,2356$-$309 with the \hess\ 
Cherenkov telescopes in 2004 is reported.
With a redshift of $z = 0.165$,
\mbox{H\,2356$-$309}
is one of the most distant AGN detected at VHE energies so far.
{
\mbox{H\,2356$-$309} was observed by \hess\
from June to December 2004 (see sections~\ref{hessobs} and~\ref{hessres}).
Simultaneous observations were carried out with RXTE {(Rossi X-ray Timing Explorer)}
in X-rays 
on 11th of November 2004 
(see section~\ref{rxtesection}), with
the Nan\c{c}ay decimetric radio telescope (NRT)
between June and October 2004
(see section~\ref{nrtsection})
and with ROTSE{-III} (see section~\ref{rotsesection})
in the optical, covering the whole 2004 \hess\ 
observation campaign. 
}

\section{\hess\ Observations} \label{hessobs}

The system of four \hess\ ACTs, located on the
Khomas Highlands in Namibia (23$^\circ$16'18''S 16$^\circ$30'00''E), 
is fully operational since December 2003.
For a review see, e.g., \citet{hinton:2004a}.
\hess\ data are taken in runs with a typical duration of 28 minutes.
The data on H\,2356$-$309 were taken with the telescopes pointing
with an offset of 0.5$^\circ$ 
relative to the object position (wobble mode,
{offset in either} right ascension {or} declination). 
The sign of the offset is alternated for successive runs to reduce systematic
effects.
H\,2356$-$309 was observed with the complete stereoscopic system 
from June to December 2004 for a total raw observation time of 
more than 80 hours.
In order to reduce systematic effects that arise due to varying
observation conditions, 
quality selection criteria are applied 
before data analysis on a run-by-run basis.
The criteria are based on the mean trigger rate (corrected 
for zenith-angle dependency), trigger rate stability, weather conditions
and hardware status.
{During the 2004 observations of H\,2356$-$309 atmospheric
conditions were not optimal ({due to} brushfires),
resulting in a dead-time corrected} high-quality data set 
of $\approx$40\,h live-time at an average zenith angle of 20$^\circ$.

These data are calibrated as described in \citet{aharonian:2004c}. 
{Thereafter,} before shower reconstruction, a standard 
image cleaning {\citep{lemoine-goumard:2005a}}
is applied to the shower images to remove night-sky background noise.
{Moreover,} in order to avoid systematic effects from shower 
images truncated by the camera edge, 
only images having a distance between their centre of gravity 
and the centre of the camera of less than 2$^\circ$ are used in the reconstruction.
Furthermore, a minimum image amplitude {(i.e., the sum of the intensities
of all pixels being part of the image)} is required for 
use in the analysis to assure a good quality reconstruction.
Previous \hess\ publications are mostly based on the standard 
analysis {\citep{aharonian:2005c}}. 
Here we present results from 
the 3D Model analysis 
which 
{is presented in detail in \citet{lemoine-goumard:2005a} and} 
was also used in \citet{aharonian:2005d}.
{This method uses} independent calibration and simulation chains
and 
{is} briefly described in the following 
{paragraphs}.


\subsection{\hess\ 3D Model Analysis}
The principle of the 3D Model reconstruction method 
\citep{lemoine-goumard:2004a,lemoine-goumard:2005a,aharonian:2005d}
is based on a 3-dimensional (3D) shower model using the stereoscopic information
from the telescopes.
The shower is modelled as a 3D Gaussian photosphere with anisotropic angular
distribution. For each camera pixel the expected 
light is calculated with a path integral
along the line of sight. 
The observed images are then compared to the model images 
using a log-likelihood fit with eight parameters,
{described in detail in \citet{lemoine-goumard:2005a}.}
For each detected shower
{at least two images} are required for the reconstruction of
the 
{angle $\theta$ 
(the angle between the object position and the reconstructed shower direction)},
{shower} core impact position {(measured as a radius with
respect to the center of the telescope array)}, energy and
the transverse standard deviation $\sigma_T$ of the shower.
{The dimensionless reduced 3D width $\omega$ 
{(used for $\gamma$-hadron separation)} is defined as 
$\omega = \sigma_T \rho / D_S$,
with the density of air 
$\rho$ and the column density $D_S$ at shower maximum.
}
The energy spectrum {of the $\gamma$-ray excess} is obtained from a comparison of 
the {reconstructed energy distributions} 
{to the expected distributions for a given spectral shape}.
For the determination of the expected number, $\gamma$-ray acceptances
calculated from simulations are taken into account. 
This \emph{forward folding} method was first
developed within the CAT collaboration \citep{piron:2000a}.
{
The quality of the reconstruction improves when only events
with a number of triggered telescopes $N_{tel} \ge 3$ are accepted.
In the analysis used in this paper, results from the 3D Model
are given using this cut. 
A spectral analysis using an $N_{tel} \ge 2$
cut yields consistent results. 
}

The on-source data are taken from a circular region of 
radius $\theta$ around the object position (on-source region). 
The background is estimated from 11 control regions (off-source regions)
of the same size and located at the same radial distance to the camera centre 
as the on-source region and normalised accordingly.
The significance {in standard deviations above background (subsequently $\sigma$)} of any excess is calculated following the likelihood method
of \citet{li:1983a}.
All cuts applied 
are summarised in Table~\ref{cuttable}.
{Cut values were optimised using simulated $\gamma$-ray samples and 
independent background data samples.}
\begin{table}[ht]\centering
 \caption{\label{cuttable}Summary of applied cuts. The cut in image 
                          amplitude is given in photoelectrons (ph.e.).}
 \begin{tabular}{ll}\hline
  cut			&	value	\\\hline
  distance		&	2\,deg		\\
  image amplitude	&	$>$60\,ph.e.	\\
  \# of telescopes	&	$\ge$\,3	\\	
  core impact position	&	$\le$\,300\,m	\\
mean reduced scaled width&	--		\\
mean reduced scaled length&	--		\\
  reduced 3D width $\omega$&	$<$0.002	\\
  $\theta^2$		&	$<$0.01\,deg$^2$\\\hline
 \end{tabular}
\end{table}

\section{\hess\ Results}\label{hessres}
\subsection{Signal}
In Figure~\ref{thetasq}, the distribution of the 
squared angular distances $\theta^2$ from H\,2356$-$309, reconstructed
with the 3D-Model ({$N_{tel} \ge 3$}), is shown.
In the data taken from the on-source region
a clear accumulation of events is seen at low $\theta^2$-values, 
{i.e. close to the position of H\,2356$-$309.}
The off-region shows a flat distribution as expected for a pure background measurement.
With a number of on-source events of \nonstr\ = \non\ 
off-source events \noffstr\ = \noff\ {and}
{a} normalisation {factor} $\alpha$=\alphafactor\ {(the ratio
between the solid angles for on- and off-source measurements)}, 
the data yield an excess of \nonstr\ - $\alpha$\noffstr\ = 
\excess\ $\gamma$-rays at a significance-level
of \significance .
A fit of a 2-dimensional Gaussian to an uncorrelated excess sky-map yields
a point-like emission and a location 
(23$h$59$m$09.42$s$\,$\pm$\,2.89$s$, -30$^\circ$37'22.7''\,$\pm$\,34.5'') 
consistent with the position of H\,2356$-$309
(23$h$59$m$07.8$s$, -30$^\circ$37'38''), 
{as obtained by \citet{falomo:1991a} using observations 
in the optical and near-infrared.}
\begin{figure}[ht]
  \resizebox{\hsize}{!}{\includegraphics{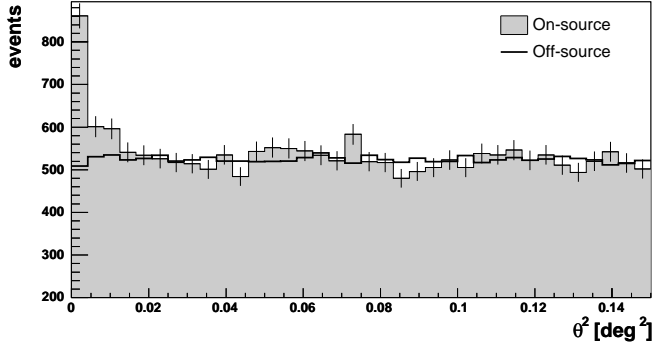}}
  \caption{Distribution of squared angular distances $\theta^2$ for
		H\,2356$-$309, using the 3D-Model reconstruction and $N_{tel} \ge 3$.}
  \label{thetasq}
\end{figure}
{These results are summarised in Table~\ref{results}.}
{Additionally, the results from the standard analysis are given.
The results for the 3D Model analysis used in this paper
are also given using $N_{tel} \ge 2$ for easier comparison with the 
standard analysis.}
\begin{table}[ht]
 \centering
 \caption{\label{results}Summary of analysis results from the 3D Model analysis
		(using two different cuts in telescope multiplicity).
		{For comparison, the standard analysis results are also
		given.} 
		}
 \begin{tabular}{llll}\hline
 				& \multicolumn{2}{c}{3D Model}	&	standard analysis	\\\hline
				& $N_{tel} \ge 3$
						&$N_{tel} \ge 2$
								& $N_{tel} \ge 2$		\\
  \nonstr\ (events)		& \non		&\nonT		&	\nonWy			\\
  \noffstr\ (events)		& \noff		&\noffT		& 	\noffWy			\\
  normalisation $\alpha$	& \alphafactor	&\alphafactor	&	\alphafactorWy 		\\
  Excess\ (events)		& \excess	&\excessT	&       \excessWy      		\\
  Significance			& \significance	&\significanceT &       \significanceWy		\\\hline
 \end{tabular}
\end{table}

\subsection{Energy Spectrum and Variability}
The differential energy spectrum obtained from the 3D-Model analysis ({$N_{tel} \ge 3$})
is shown in Figure~\ref{spectra}. The {spectral parameters 
were obtained}
from a maximum likelihood fit of a power law hypothesis 
$dN/dE = N_0 (E/\mathrm{TeV})^{-\Gamma}$ to the data,
resulting in a flux-normalisation of N$_0$ = \normalisation ,
and a spectral index of $\Gamma$ = \spectralindex.
The $\chi^2$ value of the spectral fit is 6.6 for 7 degrees of freedom, 
corresponding to a $\chi^2$-probability of $P(\,\chi^2)$ =  \chiprob.
In order to eliminate any systematic effects that might arise from poor
energy estimation at lower energies {(due to an over-estimation, on average,
of low energies), the beginning of the fit range is set to the value of the post-cuts
spectral energy threshold, i.e. 
200\,GeV for the 3D Model analysis {of this data set}.}
Systematic errors {(0.1 for the index and 20\,\% for the
flux)} are 
{dominated by atmospheric effects, i.e.}
a limited knowledge of
the atmospheric profile needed as input for the simulations.
{A detailed description of systematic errors can
be found in e.g., \citet{aharonian:2006a}}.
The parameters of the spectral fit 
are
summarised in Table~\ref{spectraltable}. {Additionally, the results from
the standard analysis are given for comparison.}
The data-points used in Figure~\ref{spectra} are listed in Table~\ref{spectralpoints}.
\begin{figure}[hbt]
  \resizebox{\hsize}{!}{\includegraphics{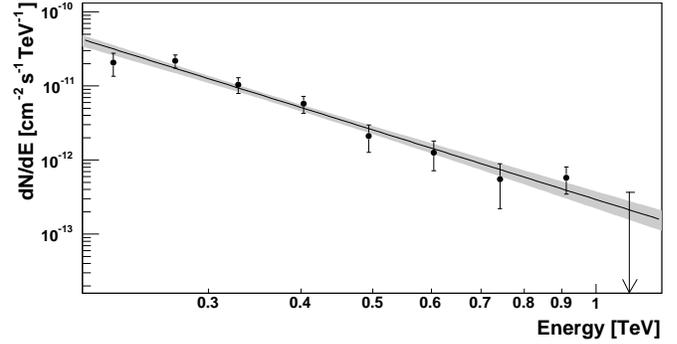}}
  \caption{\label{spectra}
		Differential energy spectrum of H\,2356$-$309, as obtained from
		the 3D Model analysis. The shaded area gives the
		confidence region {(1 standard deviation)} for the spectral 
		shape under the assumption of a power law.
		{The upper limit (arrow) 
		is given with a 99\,\% confidence level.}
	}
\end{figure}
\begin{table}[ht]
 \centering
 \setlength{\tabcolsep}{1.6mm}
 \caption{\label{spectraltable}Summary of the parameters of power law fits 
		($dN/dE = N_0 (E/\mathrm{TeV})^{-\Gamma}$).
		{For comparison with the 3D Model analysis (1) used here,
		the results from} the standard analysis (2) 
		{are also given.}
		{Both analyses yield consistent 
		results.}
		In addition to the parameter values, the $\chi^2$ 
		probabilities of the fits are given.}
 \begin{tabular}{llll}\hline
		&  $\Gamma$		& ${N_0}$	&  ${P(\,\chi^2)}$ 	\\
		& 			& $[$\funit$]$		&				\\\hline
(1) &  \spectralindex	& \normTeV	& \chiprob			\\
(2) &  \spectralindexWy	& \normWy	& \chiprobWy			\\\hline
 \end{tabular}
\end{table}

\begin{table}[ht]
 \centering
 \setlength{\tabcolsep}{0.7mm}
 \caption{\label{spectralpoints}Differential flux 
		for different energy bins.
		The upper limit is given
		for a confidence level of 99\,\%.}
 \begin{tabular}{lcc}\hline
  E    & $\Phi$ 			&$\Delta\Phi$     		\\
 $[$TeV$]$&\multicolumn{2}{c}{$[$\fu$]$}				\\\hline
  0.223& 2.06$\,\times\,10^{-11}$	& 7.13$\,\times\,10^{-12}$	\\
  0.270& 2.18$\,\times\,10^{-11}$	& 4.36$\,\times\,10^{-12}$	\\
  0.329& 1.04$\,\times\,10^{-11}$	& 2.46$\,\times\,10^{-12}$	\\
  0.403& 5.77$\,\times\,10^{-12}$	& 1.48$\,\times\,10^{-12}$	\\
  0.494& 2.12$\,\times\,10^{-12}$	& 8.51$\,\times\,10^{-13}$	\\
  0.604& 1.26$\,\times\,10^{-12}$	& 5.45$\,\times\,10^{-13}$	\\
  0.742& 5.50$\,\times\,10^{-13}$	& 3.29$\,\times\,10^{-13}$	\\
  0.912& 5.77$\,\times\,10^{-13}$	& 2.31$\,\times\,10^{-13}$	\\
  1.113& $<$4.06$\,\times\,10^{-13}$	& --\\\hline

 \end{tabular}
\end{table}
{The average integral flux above 200\,GeV in the year 2004 
({fitting} with a {fixed} spectral index of 3.09)
is $\overline{I(>200\,\mathrm{GeV})}$ = \iflux\ {(statistical error only)}.
Light-curves of I($>$200\,GeV) versus the modified Julian date (MJD) of the
observation} are shown in Figure~\ref{lc} for two different time-scales. 
The monthly flux variation is shown in the upper panel and
the average monthly flux from June to December of 2004 is shown in the lower panel.
A fit of a constant yields no evidence for nightly variability {(P(\,$\chi^2$)=0.18)}. 

{The ASM {(All Sky Monitor)} shows no significant 
{X-ray} excess nor variability in the same monthly intervals.}
\begin{figure}[ht]
  \centering
  \resizebox{\hsize}{!}{\includegraphics{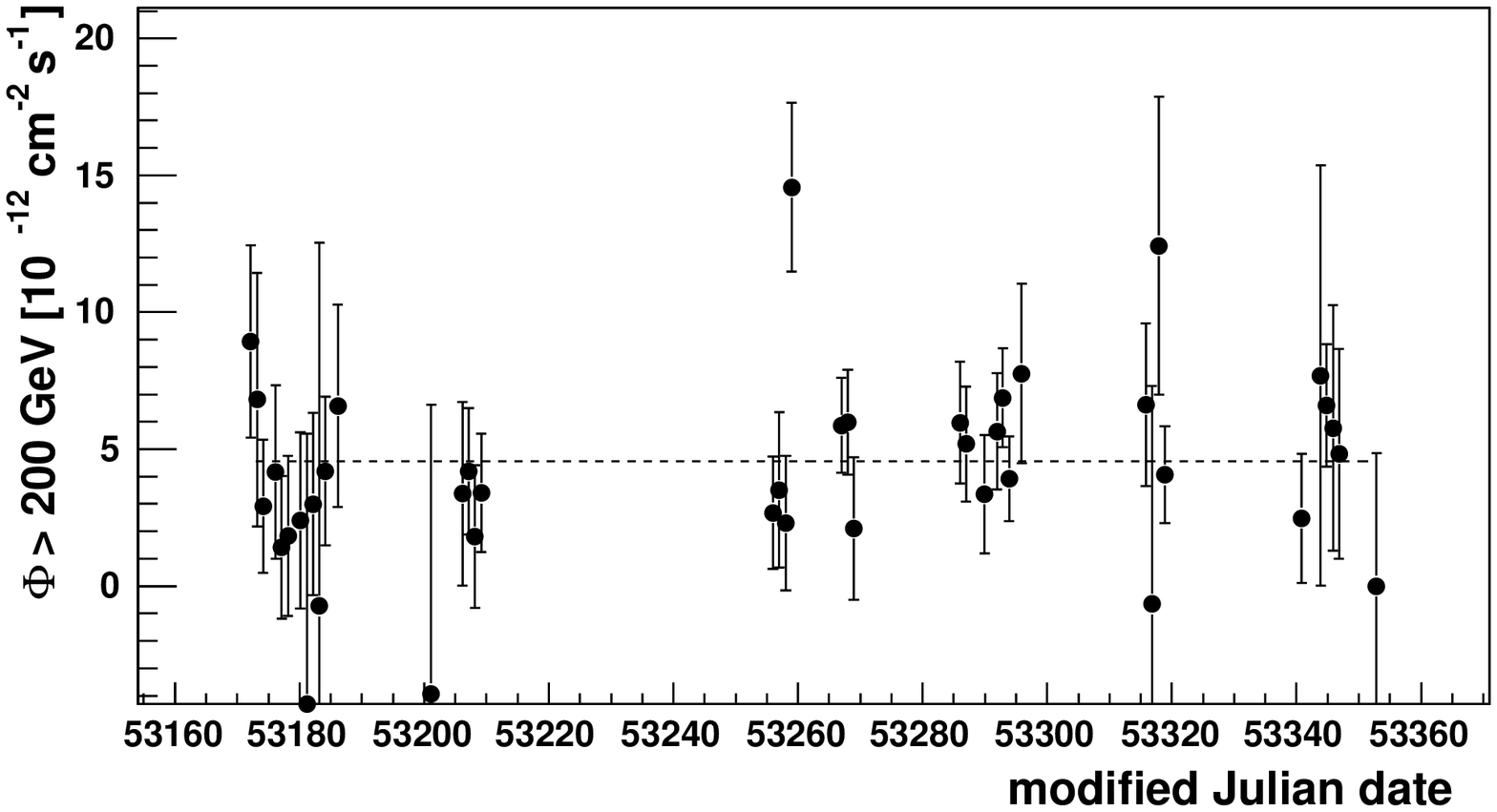}}
  \resizebox{\hsize}{!}{\includegraphics{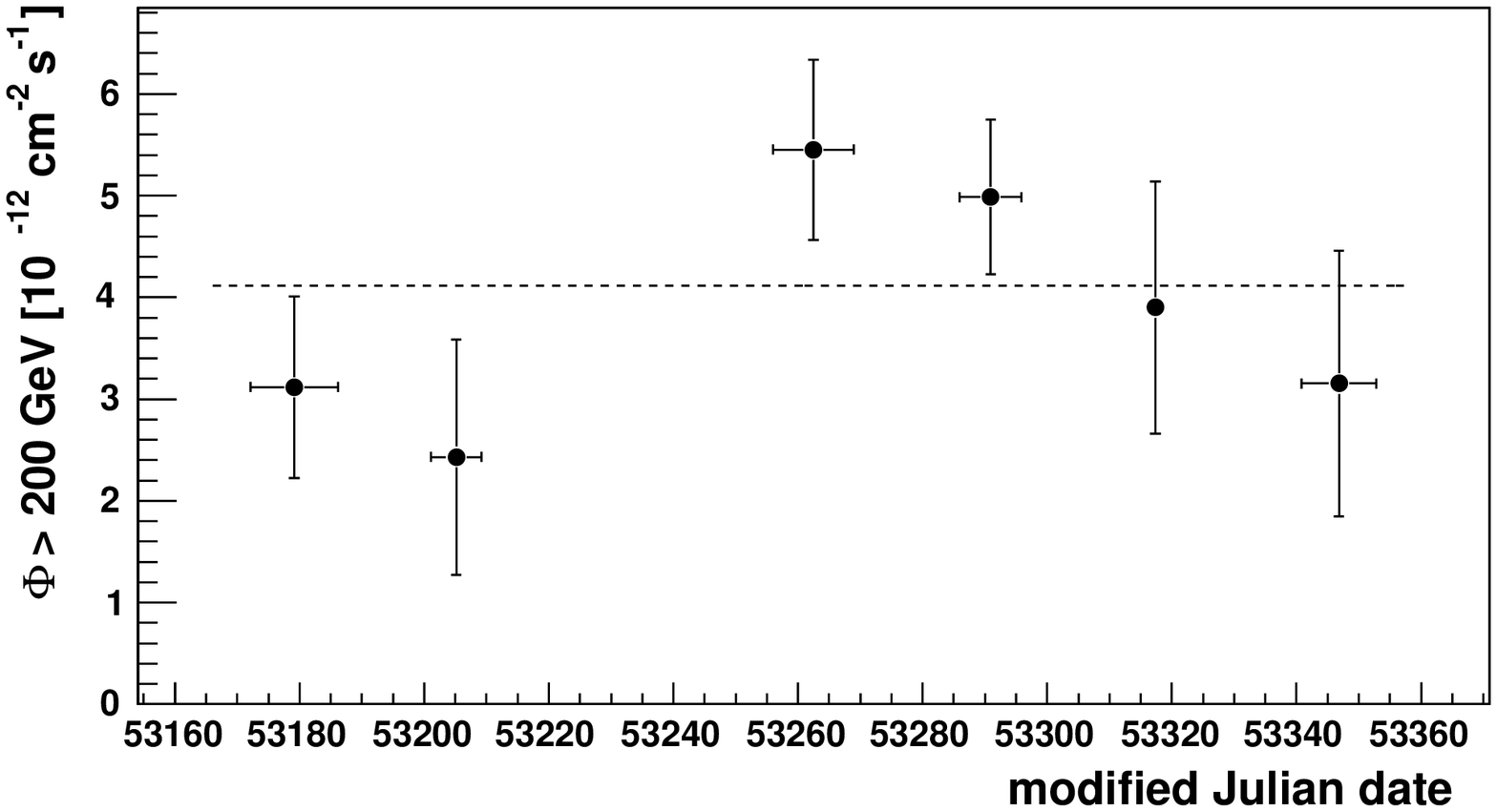}}
  \caption{Flux above 200\,GeV from H\,2356$-$309 as measured by \hess\ (3D Model Analysis) in 2004. 
		\emph{Upper panel:} each point corresponds to the flux
		measurement in one night.
		\emph{Lower panel:} averaged flux in six different
		time windows (June to December 2004).
		Both light-curves are consistent with a constant emission over
		time. 
		{The RXTE observations (see Section~\ref{rxtesection}) 
		were carried out on MJD 53320. During that night, no \hess\ observations
		could take place due to poor weather conditions.}
		}
  \label{lc}
\end{figure}

\section{Multi-Wavelength Analysis and Results}

\subsection{RXTE analysis}
\label{rxtesection}
The RXTE/PCA {\citep{jahoda:1996a}} observed H\,2356$-$309 
twice for a total of 5.4~ks on 11
November 2004 after a {Target of opportunity} was triggered on this target. 
{Due to poor weather conditions, \hess -observations were 
not possible on the night of November 11, or on the 2 prior nights.
However, the RXTE observations can be considered 
simultaneous to the \hess\ observation campaign of November.}
The STANDARD2 data were extracted using the ftools in
the HEASOFT 6.0 analysis software package provided by NASA/GSFC and
filtered using the RXTE Guest Observer Facility recommended
criteria. Only the signals from the top layer (X1L and X1R) are used
from the PCA.  The average spectrum shown in the SED (Figure~\ref{sed}) is derived by
using PCU0, PCU2 and PCU3 data. The faint-background model is used and only
the 3--20 PHA channel range is kept in {\tt XSPEC v. 11.3.2}, or
approximately 2--$10\,\rm keV$. 
The column density is fixed to the Galactic value of $N_{\rm H} =
1.33\times 10^{20}\,\rm cm^{-2}$ obtained from the PIMMS 
nH program\footnote{See http://legacy.gsfc.nasa.gov/Tools/w3pimms.html} 
and is used in an absorbed power law fit. 
This yields an X-ray photon index $\Gamma_X
= 2.43 \pm 0.11$ and a flux of $9.7 ^{+0.3}_{-1.3}\times10^{-12}\,\rm
erg\,cm^{-2}\,s^{-1}$ in the 2--$10\,\rm keV$ band. The $\chi^2$ of
the fit is 11 for 14 degrees of freedom, or a $\chi^2$ probability
$P(\,\chi^2) = 0.68$. This flux level is approximately a factor of 3
lower than what was reported from BeppoSAX observations by
\cite{costamante:2001a} and with a softer
photon index than what was observed above 2 keV. 

\subsection{NRT analysis}  \label{nrtsection}
The Nan\c{c}ay radio-telescope is a single-dish antenna with a
collecting area of $200\times34.56$ m$^2$ equivalent
to that of a $94\,{\rm m}$-diameter parabolic dish \citep{vandriel:1996a}. The
half-power beam width at $11\,\rm cm$ is $1.9\,{\rm arcmin}$ (EW)
$\times 11.5\,{\rm arcmin}$ (NS) (at zero declination), and the system
temperature is about $45\,\rm K$ in both horizontal and vertical
polarisations. The point source efficiency is $0.8\,{\rm K}\,{\rm Jy}^{-1}$,
and the
chosen filter bandwidth was 12.5 MHz for each polarisation, split
into two sub-bands of 6.25 MHz each. 
Data {were processed using} the Nan\c{c}ay
local software NAPS and SIR.

A monitoring program with this telescope on extragalactic sources
visible by both the NRT and \hess\ is in place since 2001. For the
campaign described here it consisted of a measurement at $11\,\rm cm$
every two or three days. 
Between 4 and 14 individual
1-minute drift scans were performed for each observation, and the flux
{calibration was done} using a calibrated noise diode emission for
each drift scan. 
The average flux for the measurements carried out
between 11 June and 10 October 2004 was $40\pm8\,\rm mJy$.
{This observed flux is {most likely} dominated by emission produced in jet regions
further out from the core and thus represents an upper limit of 
{any} emission model for the total SED.}

\subsection{ROTSE-III Analysis}
\label{rotsesection}
The ROTSE-III {(Robotic Optical Transient Search Experiment)} array is a world-wide network of four 0.45\,m robotic, 
automated telescopes built for fast ($\approx$ 6\,s) response
to GRB triggers from satellites such as HETE-2 (High Energy
Transient Explorer 2) and Swift. The ROTSE-III telescopes
have a wide ($1.85^\circ\times1.85^\circ$) field of view imaged onto a
Marconi 2048\,$\times$\,2048 pixel back-illuminated thinned CCD and are 
operated without filters. The ROTSE-III systems are described
in detail in \citet{akerlof:2003a}.
The ROTSE-IIIc telescope, located at the \hess\ site, has been
used to perform an automated monitoring programme of blazars, including
H\,2356$-$309. 
Data is analysed as described in \citet{aharonian:2005g} and references therein.
During the observation periods covered by \hess\ 
the apparent R-band magnitude m(R) from H\,2356$-$309
as measured by ROTSE-III has its maximum at m(R)\,=\,16.1 and its minimum at m(R)\,=\,16.9.
{The host galaxy has been resolved in the optical 
  \citep{falomo:1991a,scarpa:2000a} and near-infrared \citep{cheung:2003a}. 
  These observations show that H\,2356$-$309 is a normal
  elliptical galaxy
  with an effective radius of about 1.8\,arcsec in the R band. 
  The contribution of the galaxy to the observed
  ROTSE-III flux is estimated to be
  m(R)\,=\,17 using a standard de Vaucouleurs radial profile.}

\section{Discussion}
In Figure~\ref{sed}, a broad-band SED obtained from archival data,
and simultaneous optical (ROTSE-III) and X-ray (RXTE) data together with the 
\hess\ results presented in this paper is shown.
{Additionally, the result of a 
 simple leptonic model is given as a solid line.}
  The simplest leptonic scenario
  is a one-zone homogeneous, time independent, synchrotron
  self-Compton (SSC) model as initially proposed by
  \citet{jones:1974a} for compact non-thermal extragalactic sources.
\begin{figure*}[htb]
  \resizebox{\hsize}{!}{\includegraphics{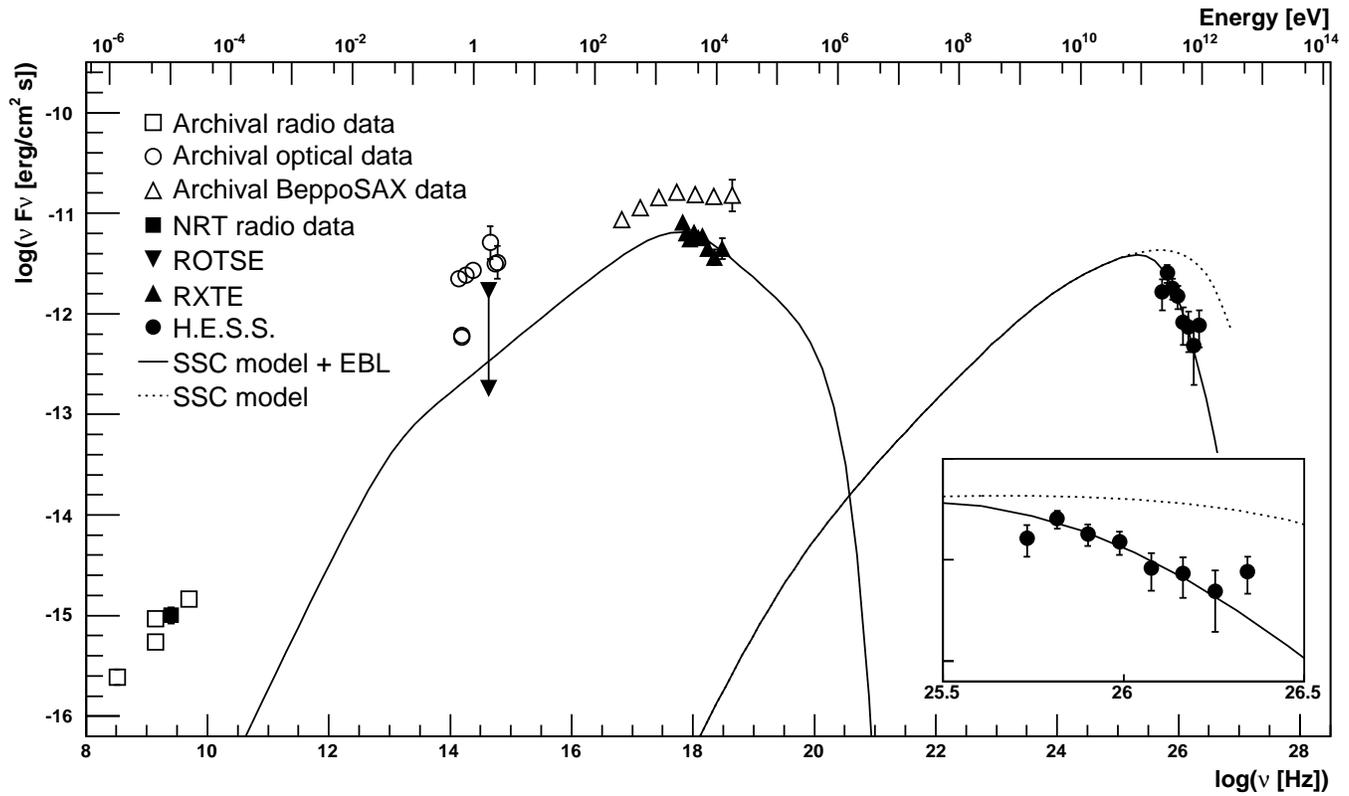}}
  \caption{Spectral energy distribution of H\,2356$-$309. 
		Above 200\,GeV, the results obtained in this paper are used.
		Data from NRT, ROTSE{-III} and RXTE are simultaneous to the \hess\ observations
		and shown as filled symbols.
		All other data are archival and shown as open symbols.
                {The ROTSE{-III} flux is given after galaxy subtraction.}
	        {Radio emission arises from regions further out in the jet.}
		A single-zone homogeneous SSC model described in the text 
		is shown as a solid line. The dashed line shows the SSC model 
		without absorption.
                The inlaid box shows a zoom in the VHE regime.
		}
  \label{sed}
\end{figure*}
{Here, we adopt a description with a spherical emitting region of}
radius $R$ and homogeneous magnetic field $B$, propagating
with Doppler factor $\delta$ with respect to the observer. 
{The high energy electron distribution is described by a broken power
law between Lorentz factors $\gamma_{\rm min}$ and $\gamma_{\rm max}$, 
with a break at $\gamma_{\rm b}$ and a normalisation $K$
\citep{katarzynski:2001a}.}
{{This} SSC scenario is used, taking into account absorption by the EBL,
to reproduce the simultaneous data of H\,2356$-$309.}
The density of the EBL is not well known in the $\mu$m wavelength regime.
Given the high redshift of H\,2356$-$309 ($z = 0.165$) and a 
comparatively hard VHE spectrum, important constraints on the EBL density 
can be derived from the \hess\ data.
This question is addressed in detail in \citet{aharonian:2005a} where
constraints on the density of the EBL are derived from \hess\ observations
of 1ES\,1101$-$232 and H\,2356$-$309.
Here, we use the P0.45 parametrisation from this paper which is very close to
the lower limit from galaxy counts.

As shown in Figure~\ref{sed},
{using a model with} a reasonable set of
parameters provides a satisfactory fit 
{to the simultaneous x-ray and VHE data}.
The emitting region is characterised by $\delta=18$, $B=0.16$\,G and
$R=3.4\,\times\,10^{15}$\,cm.
The electron {power-law} distribution is described by 
$K=1.2\,\times\,10^{4}$\,cm$^{-3}$,
$\gamma_{\rm min}=10^{3}$, $\gamma_{\rm max}=3\,\times\,10^{6}$.
The Lorentz factor at break energy $\gamma_{\rm b}$ is located
at $\gamma_{\rm b} = 2.5\,\times\,10^5$ to place
the peak emission in between optical and X-rays 
{while providing a good fit to the \hess\ data}.
We take the canonical index $\alpha_{1}$=2 for the low-energy 
end and found  $\alpha_{2}$=4.0
for the high-energy end so as to fit the {observed} X-ray power law spectrum.
Lowering $\gamma_{\rm min}$ extends the fit to lower frequencies and enhances
IC emission in the MeV-GeV domain.
Synchrotron self-absorption cuts off emission below IR frequencies
when using low values of $\gamma_{\rm min}$.  Radio emission arises
from regions further out in the jet. {Similar to the case of PKS\,2155-304
\citep{aharonian:2005g} we cannot exclude a possible contribution of
such an extended region to the optical flux measured by ROTSE{-III}.
This may soften some of the above-mentioned constraints}.

{Although our VHE observations provide strong constraints on the physical
parameters of single-zone SSC models, 
there is still some freedom of choice for the parameters 
that could be constrained {further} by 
a better understanding of the origin of the optical emission, 
a better spectral coverage in the X-ray and sub-TeV region
{and the observation of possible variability.}.}

\section{Conclusions}
The high frequency peaked BL Lac object H\,2356$-$309, located at a redshift
of $z = 0.165$, was discovered in
the VHE regime by the \hess\ Cherenkov telescopes.
Two different reconstruction and analysis methods were applied to the data
both yielding consistent results.
No strong evidence for variability in the VHE band is found 
within the \hess\ observations.
The same holds true in the X-ray band, where the object {does} not show
any strong flux variability, neither in the ASM nor in the pointed observations.
Additionally, the RXTE flux, observed simultaneously to 
the \hess\ observations, is lower than the previously-measured
BeppoSAX flux.
This might indicate that our observations took 
place during a relatively low state of emission.

For the first time, an SED comprising simultaneous radio, optical, X-ray and VHE
measurements was made.
{A simple one-zone SSC model, taking into account absorption by the EBL \citep{aharonian:2005a},
{provides} a satisfactory description of these data}.

Given the high redshift of the object, the observed \hess\ spectrum 
provides strong constraints on the density of the
EBL \citep{aharonian:2005a}.
Future observations of H\,2356$-$309 with \hess\ will improve the accuracy of
the spectral measurement and might also allow an extension of the observed
spectrum to higher energies. This will provide further constraints
on the absorption of $\gamma$-rays by the EBL.

\begin{acknowledgements}
The support of the Namibian authorities and of the University of Namibia
in facilitating the construction and operation of \hess\ is gratefully
acknowledged, as is the support by the German Ministry for Education and
Research (BMBF), the Max Planck Society, the French Ministry for Research,
the CNRS-IN2P3 and the Astroparticle Interdisciplinary Programme of the
CNRS, the U.K. Particle Physics and Astronomy Research Council (PPARC),
the IPNP of the Charles University, the South African Department of
Science and Technology and National Research Foundation, and by the
University of Namibia. We appreciate the excellent work of the technical
support staff in Berlin, Durham, Hamburg, Heidelberg, Palaiseau, Paris,
Saclay, and in Namibia in the construction and operation of the
equipment.

The authors acknowledge the support of the ROTSE-III collaboration.
Special thanks also to R. Quimby from
the University of Texas for providing tools for data-reduction.
\end{acknowledgements}

\bibliographystyle{aa}

\newcommand{\thc}{(\hess\ Collaboration) }

\end{document}